\documentclass[aps,prl,showpacs,preprint]{revtex4}

\usepackage{psfig}
\usepackage{graphics}
\usepackage{subfigure}

\newcommand{\beq}{\begin{equation}}
\newcommand{\eeq}{\end{equation}}

\newcommand{\bef}{\begin{figure}}
\newcommand{\eef}{\end{figure}}

\def\F{\mathcal{F}}
\def\nin{\noindent}

\begin{document}

\title{Nonlinear Behavior in Ferromagnetism: \\ Simple Example and 
Possible Implications}

\author{Eshel Faraggi}
\email{faraggi@physics.utexas.edu}
\affiliation{Physics Department, The University of Texas at Austin, 
Austin, 
Texas 78712}

\begin{abstract}
Two cases of a phenomenological model for ferromagnetism are considered, discrete and continuous. And the relationship, in general, between discrete
and continuous models explored. In a similar way to the logistic map behavior, the  continuous case is exactly solvable while the discrete one
contains the bifurcation route to chaos. Through the ferromagnetic interpretation I comment on the relevance of this to understanding evolution
of systems in time, the role of the configuration space in chaotic behavior, and how this understanding may lead to new exotic magnetic phenomena.
\end{abstract}


\maketitle

It has been known for some time in chaos theory that models behave in 
different ways when the dynamics is performed using different types of 
time evolution~\cite{D89,A2}. A simple example of this is the {\it 
logistic map}, for which the continues time case is fully solvable for 
all finite parameter values while for discrete time the solution ranges 
from solvable to chaotic in a route known as the {\it route to chaos}. 
In this paper I analyze another simple example of this 
dual behavior, for a model based on two essential characteristics of 
ferromagnetism: symmetry breaking of the magnetization at the critical 
point and dissipating massless dynamics. The model considered is related 
to the Landau Ginzburg model of magnetism. 
By keeping the model simple the limit of the continuous case can be 
solved exactly and since the model contains the essential 
characteristics of ferromagnetism we have some tools that we can use to 
interpret the results.

The outline of the article is as follows. The model is introduced and 
then analyzed with the assumption of continuous time evolution. For this 
case one finds an initial-condition dependent, predictable relaxation 
toward the  fixed points of the potential energy, for all range of 
parameters. Next, the basis for discrete time evolution and the 
relationship between discrete and continuous time evolution are 
discussed. Finally the discrete case is explored and the same relaxation 
behavior is found for some parameter regime but new types of behavior 
(leading to chaos) for other parameter regimes. I finish by interpreting 
this difference, that results from a different treatment of time, and 
comment on the importance of this to our understanding of the way 
processes evolve with time, and indicate systems where such behavior may 
be found.

The physical picture is as follows. Assume a bulk ferromagnet and let 
$\F(\Phi)$  be its total free energy, with $\Phi$ the total scalar  
magnetization in a given direction. If locally $\F(\Phi)$ is a proper 
physical function, it must have an appropriate expansion as a power
series. The sign of $\Phi$ is an artifact of the choice of south 
magnetic pole, so $\F( \Phi )= \F(- \Phi)$ since reversing this choice
should have no physical relevance. If we now take the fourth order 
approximation and neglect an additive term in the free energy we find, 

\beq
\label{freeE}
{\F}={ -{a \over 2} \Phi^2 + {b \over 4} \Phi^4 } 
\eeq

\nin
with $a, \, b$ the expansion coefficients. To 
ensure that infinite magnetization 
is excluded we must have $b>0$. Since $b$ is a second order parameter it 
is 
assumed for the rest that it is a constant. Equation~(\ref{freeE}) has 
two characteristic forms as we vary $a$, for $a<0$ there is a single 
potential well around $0$, we can identify this regime as paramagnetism.
In the second regime, for $a>0$, there is a double well potential. This
broken symmetry regime is identified with ferromagnetism. Indeed for
this model we can write $a \propto ({ T_c - T})$. Where $T, \, T_c$ are
the temperature and the Curie temperature respectively.

Changing the magnetization is a strongly dissipative process that
carries little to no inertia. To model this I use dissipative Newtonian
dynamics for a massless particle and arrive at the following equation of
motion for the magnetization $\Phi$,

\beq
\label{lg2}
{\dot \Phi} = M (a \Phi - b{\Phi^3}),
\eeq

where $M$ is the mobility and $1 \over M$ is related monotonically to
a damping coefficient also known as the Khalatnikov damping constant.
We can integrate this equation immediately to
find the following solution for positive initial conditions,

\beq
\Phi (t) = {\sqrt{ {e^{2 M a t}} \over { {\frac{b}{a}} {e^{2 M a t}} + |
\frac{1- {\frac{b}{a}} {\Phi_0^2}}{{\Phi_0^2}}}|}}.
\eeq

For negative initial conditions we find the solution to be $- \Phi (t)$.
For the trivial initial condition we find the trivial solution. The
non-trivial long time behavior exhibits an asymptote located at the
fixed point of the free-energy, $ \lim_{t \rightarrow \infty} \Phi(t) =
Re({\pm \sqrt{ a \over b}})$. This solution is completely deterministic
for all finite values of parameters and initial conditions, by this it is 
meant that for any desired accuracy of prediction past some time $t_0$,
one can give a requirement on the accuracy of the initial conditions at
$t_0$ which is good for all times $t>t_0$. As we shall see shortly this
situation is rather different in the discrete case.

Before going into the discrete case we should explore the relationship
between it and the continuous case. I claim there is a physical
correspondence between the discrete and the continuous
models. Usually the continuous solution would be called analytic and
appear more pleasing to the trained physical eye. But there is no
a-priori reason to assume that systems evolve continuously with time,
passing through a continuous number of states. Indeed one can imagine
how discrete processes are possible, where systems make jumps from state
to state and between jumps remain in their state for some time interval
$\Delta t$.

To relate continues and discrete models one needs a mesh. For
simplicity I assume a mesh with constant $\Delta t$, i.e. we have a discrete 
linear relationship for the magnetization,

\beq
\label{phin}
\Phi_n = {\Phi (t_0 + n \Delta t)}.
\eeq

In general an $m^{\mbox{th}}$ order differential model will be given by
the condition $f(\Phi, \, {\dot \Phi}, \, \ldots , \, \Phi^{(m)}(t))
=0$, with $m$ initial conditions. Here, $\Phi^{(n)}(t) = {{d^n \Phi}
\over {d t^n}}$. A general $m^{\mbox{th}}$ order difference model is
given by, $g_{\Delta t}(\Phi_n, \, \Phi_{n+1}, \, \ldots , \,
\Phi_{n+m}) =0$, with $m$ initial conditions.

Let $g_{\Delta t}=0$ be the difference model that one gets when
discretizing a differential model $f=0$, by approximating the
derivative ${\dot \phi} \approx 
{{\Phi_{n+1} 
- \Phi_n} \over {\Delta t}}$ and similarly \footnote{There is some 
ambiguity in the choice of the discretization of 
the differentials. Since this ambiguity involves only translation of the 
order of $\Delta t$ in time I assume that we 
can safely neglect it.} for higher derivatives. And let $C_{\Delta t}$  
the discretizing transformation. Symbolically this can be writen as 
${C_{\Delta t}(f=0)}={(g_{\Delta t} =0)}$. From analysis we know that 
$C_{\Delta t} \longrightarrow Id$ when ${\Delta t} \rightarrow 0$, where 
$Id$ is the identity. From this general property one can show that for 
small enough $\Delta t$, the $\, C_{\Delta t}$ mapping 
preserves the physical essence of the model. Stated more precisely, for 
a given mesh of time there is a unique correspondence between the 
differential and the difference models. Because we know that going from 
the difference model by taking the limit $\Delta t \longrightarrow 0$ 
one gets the corresponding differential model. If we show that this 
process is invertible, we can claim that the differential and difference 
equations of motion correspond to the same physical system with 
continuous or discrete evolution respectively. This invertiblity for 
$1^{st}$ order systems is now shown.

Assume for simplicity that a differential and a difference model are 
given by ${\dot \Phi} = f(\Phi), \; \Phi_{n+1} = g(\Phi_n)$ 
respectively, where $f, \; g$ are well behaved functions. For a given 
mesh size  ${\Delta t} >0$, let $\Phi_n$ 
be given by equation~(\ref{phin}) and let ${C_{\Delta t}({\dot
\Phi}=f(\Phi))}=({{{\Phi_{n+1} - \Phi_n} \over {\Delta t}}=f(\Phi_n)})$ 
be the correspondence between differential and difference equations. Now 
if,
 
$${C_{\Delta t}({\dot \Phi}=f_1(\Phi))} = {C_{\Delta t}({\dot 
\Phi}=f_2(\Phi))}$$

\nin
it is easy to see that $f_1=f_2$, just by allowing the initial conditions
to vary continuously, i.e. $C_{\Delta t}$ is 1-1.  To show onto, we can
take $f(\Phi) = \frac{1}{\Delta t} [g(\Phi) - \Phi]$ where 
$g$ is the difference model. Then, 

$$C_{\Delta t} ({\dot \Phi} = f(\Phi)) = (\Phi_{n+1} = g(\Phi_n)).$$

This shows that, for a given mesh of time, for every difference model 
there exists a unique differential model and vice versa. And thus, any 
changes that a given model will exhibit when going from continuous to 
discrete time will result directly from this discretization and not from 
any other change to the underlying physical dynamics.

I now show the discrete case for this model of magnetism. If we 
discretize equation~(\ref{lg2}) we find the following difference
equation for the magnetization,

\beq
\label{dislg2}
\Phi_{n+1} = g(\Phi_n) = \Phi_n + c_1[\Phi_n - c_2 (\Phi_n)^3]
\eeq

\nin
where $c_1 = M a {\Delta t}$ and $c_2 = {b \over a}$. $\Delta t$ is a 
finite 
time element.

The fixed points for this map, given by the condition $\Phi_{n+1} =
\Phi_n$ , are  $ \Phi^* = 0$ or $\Phi^* = \pm
{1 \over \sqrt{c_2}}$. To understand the nature of these fixed points 
their stability is analyzed,

$$ g ' (\Phi^*) = 1 + c_1 [1 - 3 c_2 (\Phi^*)^2].$$

Or,

$$ g'(0) = 1 + c_1, $$

$$ g'({\pm \sqrt{ 1 \over c_2}}) = 1 - 2 c_1.$$

Thus for $ -2 < c_1 < 0$ the $\phi^* = 0$ fixed point is stable which 
corresponds to the paramagnetic state. For $0 < c_1 < 1$, the
$\Phi^* =  \pm {\sqrt{ 1 \over c_2}}$ is a stable fixed point for 
positive and 
negative initial  conditions respectively. We recognize this as the 
ferromagnetic state. For larger values of $c_1$ the fixes points become 
unstable and we find higher period fixed points. If we set $c_1$ even 
larger we find a dense set of unstable periodic orbits with total 
measure zero~\cite{D89}. This last scenario is known as 
chaos and the route we have taken is the period doubling route to chaos.

Remembering that $c_1 = M a {\Delta t}, \quad M > 0$ and that for a 
forward evolving system $\Delta t > 0$, we can conclude again that $a$ 
is the parameter that determines whether the system is a paramagnet or a 
ferromagnet. Note that for a backward evolving system $(\Delta t < 0)$ 
the roles of paramagnetism and ferromagnetism change and we get the 
transition from ferromagnetism to paramagnetism as we lower the 
temperature. 

The standard conclusion derived from such a treatment is that as long as we keep 
$ \Delta t$ small enough such that $c_1$ is in the stable region $(-2,1)$ we can 
analyze equation~(\ref{lg2}) with equation~(\ref{dislg2}). But there is 
some thing more fundamental that we can conclude: {\bf the nature of 
time evolution - whether discrete or continuous, leads to different 
observation in physical systems.}

\begin{figure}
\includegraphics{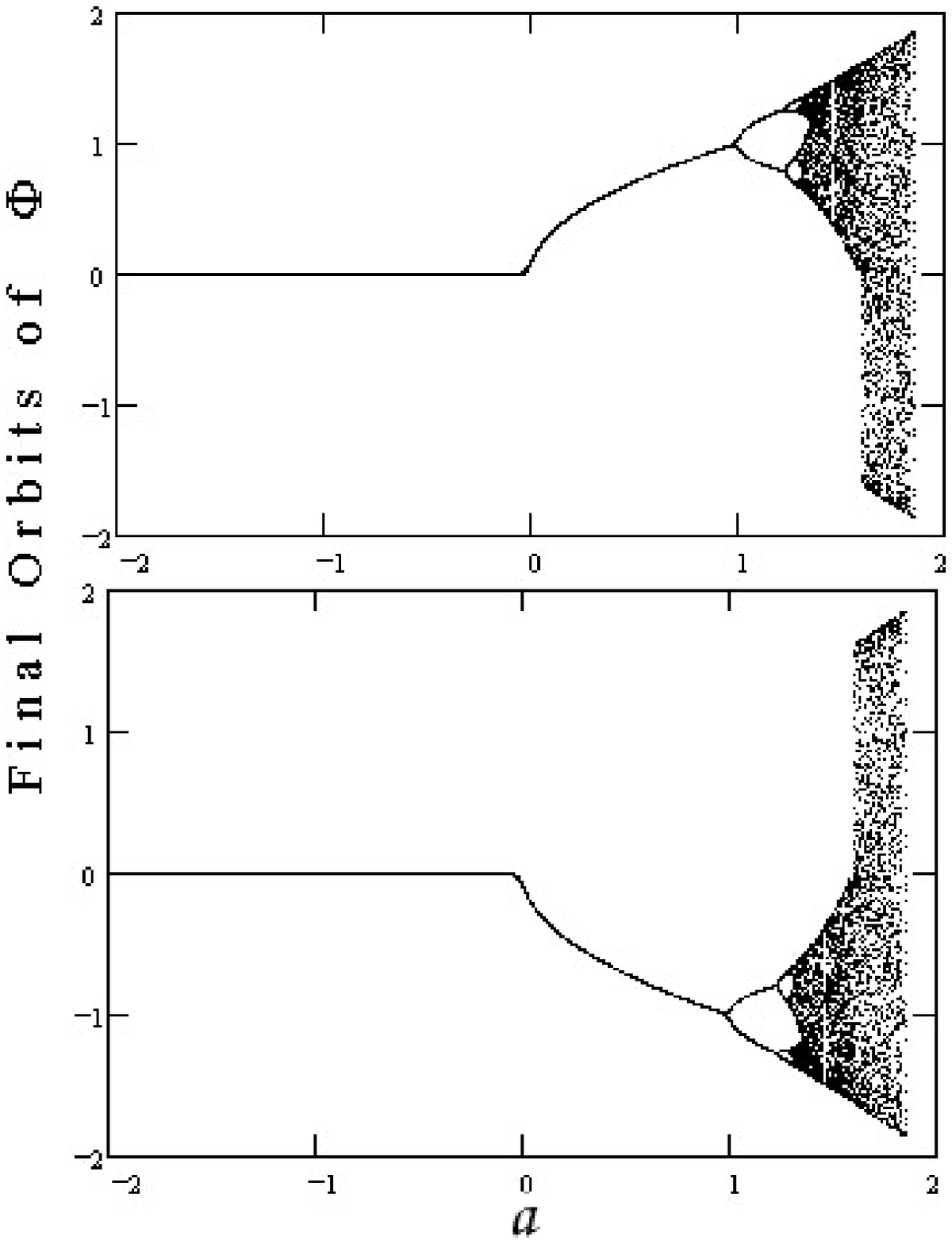}
\caption{Temperature bifurcation diagram with constant mobility for the
total magnetization in the discretized Ginzburg-Landau model, $M \Delta t = 1$ and 
$b=1$.
Top: Positive initial condition Bottom: Negative initial condition.}
\label{tbif}
\end{figure}

With this last conclusion in mind we are led quite directly to inferring 
from observations on the nature of ferromagnetism under discrete time 
evolution. For the system presented we consider two control parameters, 
the temperature and the mobility. Figure~\ref{tbif} is the bifurcation 
diagram as we vary the temperature, i.e. for a given temperature we 
iterate equation~(\ref{dislg2}) many time steps (200) and we plot the final states
it settles to (100 points). We start in the paramagnetic state with zero magnetization 
then at $T_c$ ($a=0$) we have a transition to the ferromagnetic state. As we keep
increasing $a$ we have more transitions until we reach the chaotic 
state. So, if time is discrete then we might expect that if we lower the 
temperature enough we will get a transition from the ferromagnetic state 
to a state for which the magnetization oscillates between two adjacent 
magnetization values. If we lower the temperature even further the 
magnetization difference between these two states will increase. 
However, it could turn out that time evolution is discrete 
and we would not observe such exotic magnetic states. The reason is that 
there is a physical bound on the value of $a$, set by the minimum 
possible temperature, $T=0\, K$. It could turn out, especially if we 
expect a small $\Delta t$, that in most cases the corresponding value of 
$c_1$ is well within the familiar ferromagnetic region.

\begin{figure}
\includegraphics{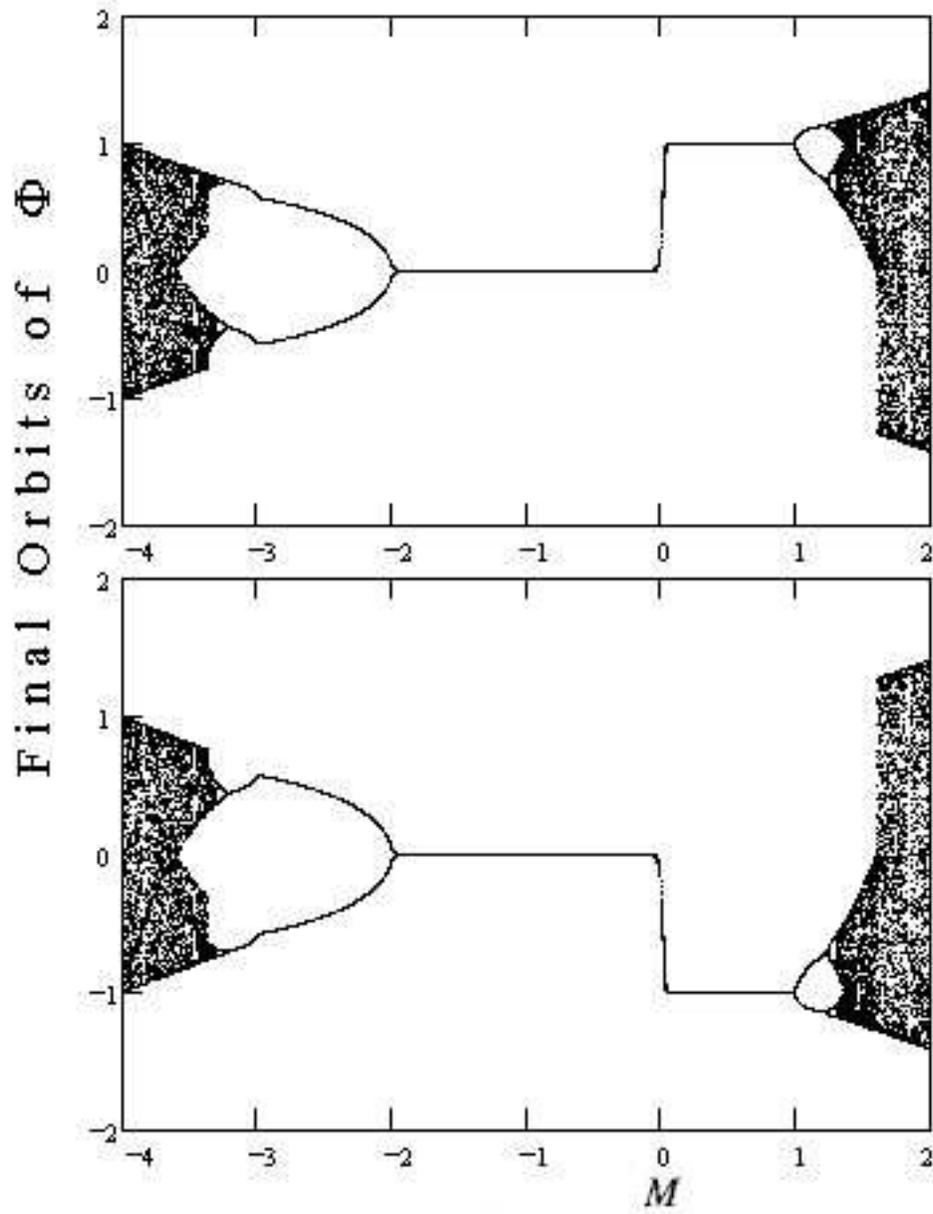}
\caption{Mobility bifurcation diagram with constant temperature for the
total magnetization in the discretized Ginzburg-Landau model, $a \Delta t =1$ and $c_2 = 1$. Top: Positive initial condition. Bottom: Negative initial condition.}
\label{mbif}
\end{figure}

The second control parameter is the mobility. Figure~\ref{mbif} is the
bifurcation diagram keeping $a, \, b$ fixed, positive and varying the 
mobility. I remark that the part of the diagram for negative values of 
the mobility was left in mainly for esthetic reasons. As we increase the 
mobility we observe much the same bifurcation diagram as before. Indeed 
it is seen that there is a value of $M$ for which the familiar 
ferromagnetic state ends and we start a new phase for which the 
magnetization oscillates between nearby
states. Moreover there seems to be no natural cutoff for the mobility 
and detection of these exotic magnetic states seems more 
plausible~\cite{A1}. Another system where detection of 
exotic states may be possible is discussed in~\cite{A2}.

In the frame-work of magnetization these nonlinearities are intriguing 
and could also provide evidence about the nature of time evolution. 
Indeed as this and other work suggest, bifurcation routes and chaos rise 
naturally from discrete dynamics, and disappear once the continuous 
limit is taken. Interestingly enough, experimental and numerical work on 
a driven magnetic system do exhibit a period-doubling route to  
chaos~\cite{A1,A3}. Another possible system where the exotic nature of 
the magnetization may be found is in a super-critical ferromagnet where 
the critical temperature (and thus $a$) is very large compared to everyday 
ferromagnets. In such systems the distance between nearest neighbor 
magnetic moments is small such that the overlap between adjacent 
wave functions and hence the exchange integral are greatly increased. An 
example of such a system is a neutron star, where the distances between 
adjacent magnetic moments are of nuclear order. Such stars exhibit a 
large-scale ferromagnetic ordering at very high temperatures which
gives a very high critical temperature. When such a star
cools down $a$ becomes large and the possibility of exotic states arises
in the realm of discrete time. Such exotic states will oscillate between
different magnetization states with period of the order $\Delta t$. If
our instruments make observations over a time much larger than $\Delta t$
we would observe only the average magnetization.

In conclusion, a simple model for the bulk magnetic state has been
studied. It is found that under continuous time evolution assumption
this model admits only a steady-state analytical solution that is fully
deterministic for all model-parameters values. Under discrete time
evolution assumption the model exhibits the same analytical solution for
some parameter values but as part of the period doubling route to chaos. This different
behavior resulted from the different assumption with regard to time
evolution and thus such observation can aid in understanding
the nature of time and of time evolution.

\begin{acknowledgments}
The author would like to acknowledge the support generously provided by
Natali J. Teszler.
\end{acknowledgments}

\clearpage

\end{document}